\documentclass[3p,review,sort&compress]{elsarticle}
\usepackage{amsmath}
\usepackage{amssymb}
\usepackage{amsfonts}
\usepackage{url}
\usepackage{hyperref}
\usepackage{epsfig}
\usepackage{graphicx}
\usepackage{multirow}
\usepackage{caption}
\usepackage{cite}
\usepackage{float}

\journal{ArXiv}

\begin{document}

\begin{frontmatter}

\title{Trends in the E-commerce and in the Traditional Retail Sectors\\During the Covid-19 Pandemic: an Evolutionary Game Approach}

\author[PUC]{Andr\'e Barreira da Silva Rocha\corref{cor1}}
\cortext[cor1]{corresponding author}
\ead{andre-rocha@puc-rio.br}
\author[PUC]{Matheus Oliveira Meirim}
\author[PUC]{Lara Corr\^ea Nogueira}
\address[PUC]{Department of Industrial Engineering, Pontifical Catholic University of Rio de Janeiro,\\ Rua Marqu\^es de S\~ao Vicente, 225, G\'avea, CEP22451-900, Rio de Janeiro, RJ, Brazil.}

\begin{abstract}
An evolutionary game model is developed to study the interplay between consumers and producers when trade takes place on an e-commerce marketplace. The type of delivery service available and consumers' taste are particularly important regarding both game payoffs and players' strategies. The game payoff matrix is then adapted to analyse the different trading patterns that were developed during the COVID-19 pandemic in both the traditional retail and e-commerce sectors. In contrast to the former, investment in logistics and warehouses in the e-commerce sector allowed for the emergence of a trend in which fast delivery and eager consumers are becoming the norm. 
\end{abstract}

\begin{keyword}
Evolutionary game theory\sep replicator dynamics\sep e-commerce\sep logistics. 
\end{keyword}

\end{frontmatter}

\section{Introduction}
\label{sec:intro}
Despite mass vaccination becoming a reality all around the world, the COVID-19 pandemic still imposes important restrictions in the economy. The pandemic not only left an impressive death tool but also changed the face of high streets for good all around the world. In the UK, well-established names disappeared from the town centres and shopping malls. Among those are Arcadia and Debenhams. The latter was a department store chain with operations dating back to the 18$^{th}$ century. After falling into administration, it was bought by an online retailer in January 2021 such that the brand will survive only on the internet (Partridge, 2021). Such trend in the traditional retail sector was also observed in Brazil. In contrast, the growth rate of the e-commerce sector in Brazil was 68\% in 2020. Online marketplaces played an important role, representing a share of 78\% of the e-commerce sector during the pandemic (Mansano, 2021). Among well-known players, Amazon Brazil faced an increase of 63\% in the number of online searches over the period Feb/20-Mar/21. Even with high street shops reopening, e-commerce is still growing, showing that the pandemic accelerated a change in consumers' behaviour.

According to Colla and Lapoule (2012), e-commerce became popular among consumers willing to reduce time spent in the purchasing process. Even before the pandemic, the growth rate of e-commerce in Brazil recorded in 2019 an increase of 22.5\% in the number of purchasing orders when compared to 2018 (Fernandes, 2020). In such growing and competitive sector, Hu et al. (2016) emphasize the importance to offer an efficient and high quality delivery service in order to satisfy and keep the existing customers loyalty as well as attract new ones. Meeting delivery schedule at low enough costs plays an important role in the achievement of such quality standard (\"{U}lk\"{u} and Bookbinder, 2012). In this context, there was an important increase in the number of firms providing logistic services (Du et al., 2018) and at the same time putting a strong effort into minimizing the time for delivery (Grondys et al., 2016).

In light of this evolving competitive environment, we employ an evolutionary game model to study the interplay between consumers and producers when trade takes place on an e-commerce marketplace. The type of delivery services available and consumers' taste are particularly important in our model setup. We then extend our analysis in order to understand the different trends that emerged in both the e-commerce and traditional retail sectors during the first year of the COVID-19 pandemic. Our results show that, while in the e-commerce sector massive investment in delivery logistics and warehouses allowed for the establishment of a trend in which fast delivery and eager consumers are the norm, in the traditional retail sector, high street shops remained in a sort of non-technological trap in which traditional street shoppers and inefficient companies are still an important share of the market.

The use of an evolutionary game approach stems from the fact that the latter has an advantage over classic games in the sense that the game is dynamic by default, best suiting a market in which firms and consumers repeatedly interact over time through trade. Moreover, we can relax the assumption of fully rational agents. Instead, we assume boundedly rational agents such that each firm's and each consumer's strategic decision is naive regarding profit maximization and is purely based on corporate culture and consumer taste, respectively. In contrast to classic game theory, agents do not necessarily play a best response. As discussed in Crawford (2013), agents' bounded rationality implies that players follow simple adaptive rules that may converge to a game equilibrium. Such kind of bounded rationality and the possibility for agents' adaptation and revision of strategy over time is implicit in the replicator dynamics governing the evolution of firms and consumers over time. Best performing strategies obtain more profit (or utility), which is reflected in those strategies having a higher growth rate over time in the replicator dynamics equations when compared to poor performing strategies. The latter grow at a lower rate up to becoming (relatively) extinct in their population, i.e., the share of individuals adopting them becomes zero. Thus, there is an implicit natural selection mechanism in replicator dynamics, with nature (instead of agents) rationally selecting the best performing strategies.

As Brian Arthur (1994) points out, the type of rationality assumed in economics demands much of human behaviour and breaks down under complicated problems due to two main reasons: (i) beyond a certain level of complexity, human logical capacity ceases to cope; (ii) in interactive situations of complication, agents cannot rely upon the other agents to behave rationally, thus being forced to guess their behaviour. In such situations, psychologists tend to agree that humans think inductively with bounded rationality, simplifying the problem. Evolutionary games have been widely used in the literature in the field of operations research applied to firm and consumer competition. In Xiao and Yu (2006) supply chain model, agents rationally maximized their utilities but they were naive regarding which preference function they should follow as the best one in the long run. Anastasopoulos and Anastasopoulos (2012) point out that evolutionary games also have advantages when compared to their classic game theory counterparts because the former are capable of explaining how players achieve the equilibrium in the long run.

The remainder of the paper is organized as follows: in Section 2, we setup the evolutionary game model. Section 3 solves the game, discussing its dynamic stability and equilibria. Section 4 applies the model to both the traditional retail and e-commerce sectors.
Section 5 concludes.

\section{Model}
\label{sec:model}
We model the interplay between buyers and sellers using an online market platform for the sale of a final good. On the demand side, there is a very large population of consumers (buyers) composed of two types of agents. On the one hand, we have a so-called eager consumer (strategy $E$) for whom time is a key driver regarding consumption. Alongside this type of consumer, we have consumers who are happy to wait longer (strategy $L$) for the delivery of their purchased goods. The proportions of each type of consumer in the population are $x$ and $1-x$, respectively. On the supply side, a very large population of producers (sellers) can also be split into a two-agent category: those who find it important to have their goods delivered as fast as possible to their consumers (strategy $F$), and producers who do not have delivery as a key strategy in their business (strategy $S$). Their proportions in the supply side of the market are $y$ and $1-y$, respectively. 

The final good delivery network from the producer warehouse to the consumer home is provided by either a courier service or the regular post. In the former case, we assume delivery time is faster and always met but requires an extra freight cost $C_f$, which in our model is paid by the seller. Delivery through the regular postal service takes longer and has no extra cost to either the seller or the buyer. Whenever a trade between a consumer and a producer takes place in our e-commerce marketplace, it can be modelled using a game theory payoff matrix. In each potential online trade, there is some probability that an $E$-type (resp. $L$-type) consumer $i$ searches and finds an $F$-type (resp. $S$-type) producer $j$. Possible meetings are given by the four pure strategy profiles $(i,j)\equiv\left\lbrace (E,F),(E,S),(L,F),(L,S)\right\rbrace $, with corresponding probabilities $p_{(i,j)}$ given respectively by $xy$, $x(1-y)$, $(1-x)y$ and $(1-x)(1-y)$, such that $\sum p_{(i,j)}=1$.

Similar to perfect competition, the final good provided by each producer is homogeneous and the amount produced by each firm is small relatively to the market aggregate supply, thus firms are price takers and sell their final good charging a market price $p$. Moreover, production technology adopted by the producers is the same across the market, thus all producers face the same marginal cost $\text{CMg}$. Product differentiation comes from the two different delivery options. Firms delivering through the courier service charge an extra amount $\delta$. In the real world, such practice is similar to the so-called FULL delivery option in the Mercado Livre online platform, in which prices are slightly higher but delivery is generally carried out within 24-hours after the purchase takes place. On average, whenever a trade takes place, a consumer buys only one unit of the final good. There is no unplanned inventory, i.e., aggregate supply equals aggregate demand and, on average, a unit of the good is only produced/delivered whenever a consumer makes a purchase. Fixed costs are zero. 

Whenever a consumer carries out a particular online search, he is paired with a producer and a potential trade might take place. The latter is successful only if an $E$-type consumer is paired with an $F$-type producer or an $L$-type consumer is paired with an $S$-type producer. In the former case, the eager consumer is happy to pay the premium $\delta$ over the regular market price $p$ in order to have his good delivered faster. If he was to meet with an $S$-type producer, he would prefer not purchasing and would carry out another search. In contrast, the $L$-type consumer only purchases if the online search results in finding a producer delivering the good through the post, thus avoiding the premium $\delta$. Thus, consumers face the following possible utilities from any particular internet search: $U_{(E,F)}>U_{(E,S)}$ and $U_{(L,S)}>U_{(L,F)}$. We also assume that if the final good is given (resp. not given) to a consumer, an eager consumer feels happier (resp. sadder) than a non-eager consumer, i.e., $U_{(E,F)}>U_{(L,S)}$ $\left(\text{resp. }U_{(L,F)}>U_{(E,S)}\right)$. Thus, $U_{(E,F)}>U_{(L,S)}>U_{(L,F)}>U_{(E,S)}$. 

Regarding the producers, each seller pays the same cost for using the e-commerce platform, thus without loss of generality, we normalize such payment to zero. Hence, a producer's profit $\Pi_T;\ T=\left\lbrace F,S\right\rbrace$ is given by $\Pi_T=\psi+\mu$; $\mu=0\text{ if type }T=S$, where $\psi=p-\text{CMg}$ and $\mu=\delta-C_f$ are respectively the operating and the logistics markups that each producer makes whenever a trade takes place. The operating markup is the profit per unit of the good produced and sold while the logistics markup is the profit per unit of the good produced and delivered. In the real world, the operating markup is generally larger than the logistics markup given the former is related to all the firm's operations while the latter is specific of a particular service provided by the firm. Hence, in our model we assume $\mid\psi\mid>\mid\mu\mid$. The game payoff matrix, with row (resp. column) players representing a consumer (resp. producer) is: 
\begin{equation}
	\bordermatrix {&F&S\cr
		E &U_1,\psi+\mu& U_4,0 \cr
		L &U_3,0& U_2,\psi\cr} 
	\label{Matrix-I}
\end{equation} 
where $U_1>U_2>U_3>U_4$, as discussed above. The payoff matrix displays the players' conflict: if on the one hand, while facing an $F$-type producer, an $E$- type consumer would be better-off, on the other hand, whenever facing an $S$-type producer, an $L$-type consumer would be better-off. The same analysis holds true from the producer's viewpoint given that a particular type of producer would be better off depending on which type of consumer is faced during a potential online trade.

\section{Results}
\label{sec:results}
In this Section, we employ an evolutionary game approach in order to solve the players' conflict and to address the evolution of both populations of consumers and producers over time. Such an approach has at least three benefits over the classic game theory approach: (i) the game in this paper models a market which is naturally a dynamic setup in the real world, in the sense that agents meet continuously over time. The system of non-linear differential equations used in evolutionary game theory is more suitable to meet such dynamic setup than the classic game theory static setup; (ii) we do not need to rely on the concept of agents' rationality. Instead, we assume the players display bounded rationality, i.e., while players are rational and fully able to compute their individual payoffs in the payoff matrix, the strategy selection process is not based on rationality. Instead of always picking the best reply strategy as recommended in classic game theory, strategy choice in evolutionary game theory relies not on rationality, but on players' behavioural traits such as corporate culture (sellers) and consumption tastes (consumers). Moreover, if over time, a given strategy tends to be more successful regarding profit or utility, players are able to adapt and might review and update/change their minds regarding which consumption/production strategy they should employ; (iii) we do not need to rely on the concept of Nash equilibrium. The game in (\ref{Matrix-I}) might display multiple Nash equilibria and, without further assumptions, classic game theory does not answer which of the equilibria will be played. Evolutionary game theory overcomes such limitation. Whenever multiple evolutionary equilibria exist, one can always anticipate which equilibrium will be played in the long run.

In order to address all three points above, we employ a replicator dynamics (RD) non-linear system of ordinary differential equations (ODE) to study the evolution of the shares of each type of player in both populations. In evolutionary game theory, RD models natural selection of strategies. Thus, more successful strategies increase their share of adoption among agents over time and less successful strategies tend to extinction and to disappear in the population in the long run. Given that $x\in\left[0,1 \right] $ (resp. $y\in\left[0,1 \right]$) is the proportion of agents adopting strategy $E$ (resp. $F$) in the population of consumers (resp. producers), the RD system based on (\ref{Matrix-I}) is:
\begin{eqnarray}
	\frac{\partial x}{\partial t}=\dot{x}&=&x(1-x)\left[  \bar{\pi}_E-\bar{\pi}_L\right]\nonumber\\ &=&x(1-x)\left[ (U_1-U_3+U_2-U_4)y+(U_4-U_2)\right] 
	\label{RDp}\\
	\frac{\partial y}{\partial t}=\dot{y}&=&y(1-y)\left[ \bar{\pi}_F -\bar{\pi}_S\right] \nonumber\\ &=&y(1-y)\left[ (\mu+2\psi)x-\psi\right]
	\label{RDq}
\end{eqnarray}	
where $\bar{\pi}_z$ is the expected profit/utility from playing strategy $z$. The stationary points $(x^*,y^*)\therefore\left\lbrace \dot{x}=0\wedge\dot{y}=0\right\rbrace$ of the system given by equations (\ref{RDp}) and (\ref{RDq}) are the corner points $(0,0)$, $(1,0)$, $(0,1)$, $(1,1)$ and the interior point corresponding to the Nash equilibrium in mixed strategies of the classic version of the evolutionary game $\left(x^*=\frac{\psi}{\mu+2\psi},y^*=\frac{U_2-U_4}{U_1-U_3+U_2-U_4}\right)$. An evolutionary equilibrium is an asymptotically stable stationary point of the RD system of ODE. In order to analyse the stability of the system at the stationary points, the non-linear system is linearised and has the following Jacobian matrix $\Omega(x^*,y^*)=\Omega^*$:
\begin{equation}
	\Omega^*= \begin{bmatrix}
		(1-2x)\left[ (U_1-U_3+U_2-U_4)y+(U_4-U_2)\right] & (U_1-U_3+U_2-U_4)(x-x^2)    \\[0.3em]
		(\mu+2\psi)(y-y^2)& (1-2y)\left[ (\mu+2\psi)x-\psi\right]
	\end{bmatrix}
	\label{Jacobiana}
\end{equation}  
The eigenvalues $\lambda$ of the Jacobian matrix are given by the roots of the equation resulting from setting the determinant $\det(\Omega^*-\lambda I)$ equal to zero, where $I$ is a 2x2 identity matrix. Given that at the corner stationary points, $x$ and $y$ are always either 0 or 1, when computed for those points, the Jacobian matrix $\Omega^*$ has non-zero entries only along the main diagonal. The eigenvalues are thus given by $\lambda_1=\frac{\partial \dot{x}}{\partial x}$ and $\lambda_2=\frac{\partial \dot{y}}{\partial y}$. The condition for asymptotic stability of the linearised system in the neighbourhood of a particular stationary point is that both eigenvalues of $\Omega^*$ have negative real parts. For the interior stationary point, the eigenvalues are given by $\lambda_{1,2}=\pm\sqrt{\frac{\partial \dot{x}}{\partial y}\cdot\frac{\partial \dot{y}}{\partial x}}$. From (\ref{Jacobiana}), $\lambda_{1,2}= \pm\sqrt{x(1-x)y(1-y)(\mu+2\psi)(U_1-U_3+U_2-U_4)}$. Given that $U_1>U_2>U_3>U_4$ and $\mid\psi\mid>\mid\mu\mid$, the phase space topology depends on the sign of the operating markup $\psi$. 

\section{Discussion}
\label{sec:discussion}
In this Section, we analyse and discuss two key cases during the COVID-19 pandemic.
\subsection{Traditional Retail Sector}
\label{subsec:1}
The first case applies the evolutionary framework to the context of traditional retail business. During the main waves of the COVID-19 pandemic, non-essential high street shops had to temporarily shut down in order to comply with restrictive measures regarding the circulation of people in the streets. Such a trend was observed in many countries around the world. 

Taking into account that such shops had already accumulated a pre-pandemic inventory, one might model such restrictive period as a market in which there was no supply shock but instead a demand shock. Sellers in the traditional retail sector faced a large negative impact in market demand due to social isolation and quarantine measures. There was excess supply of goods which could not be immediately traded, leading to a negative impact in firms' revenue due to a fall in traded volumes, while costs (existing inventory, wages, rents, expiration of perishable goods, etc.) were still being financially accrued. 

Using a standard microeconomics model of market supply and demand for market price determination, such an impact in the firms' revenue described above is similar to a fall in the market price when the aggregate supply curve remains unchanged but the aggregate demand curve faces a negative shock. Losses take place due to price falling below marginal cost, i.e., operating margin $\psi$ becomes negative in the payoff matrix (\ref{Matrix-I}). Producers would be better-off leaving the market. While such trend was adopted by important retailers around the globe, the majority of retailers braced for hard times. Within our bounded rationality framework, producers remain in the market.

We adapt the payoff matrix developed for e-commerce in (\ref{Matrix-I}) in order to apply it in the context of traditional retail. We assume consumers had already purchased the final good and, despite the latter being produced and available in the shop, a lockdown completely prevented in store collection of the good. In order to partially overcome the problem, in the case of traditional high street shops, a common observed practice adopted by some retailers was to display some phone or whatsapp number on the shop door such that buyers could order by phone and collect the good at the shop at some pre-arranged time. We assume that an $F$-seller is a high street shop adopting such practice while the $S$-seller simply remains with the shop closed and no collection arrangement is available at all. The $E$-buyer is a consumer who is eager to receive the good and is happy to arrange in store collection on the phone, while the $L$-buyer is a consumer who is happy if he can wait for the end of the quarantine period before leaving the house. 

\begin{figure}[H]
	\centering
	\begin{tabular}{cc}
		\epsfig{file=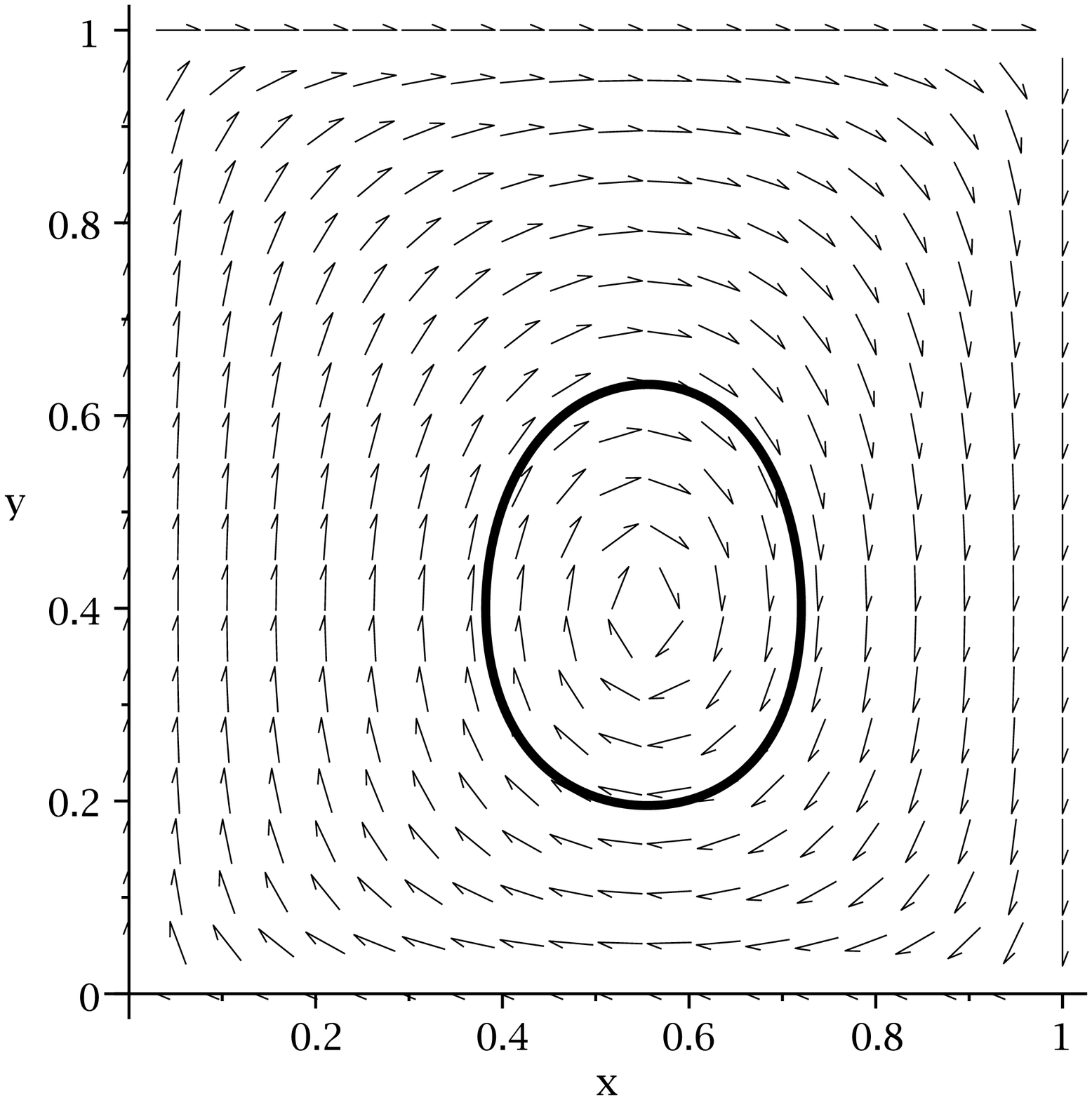,height=6cm,width=6cm,angle=0}&
		\epsfig{file=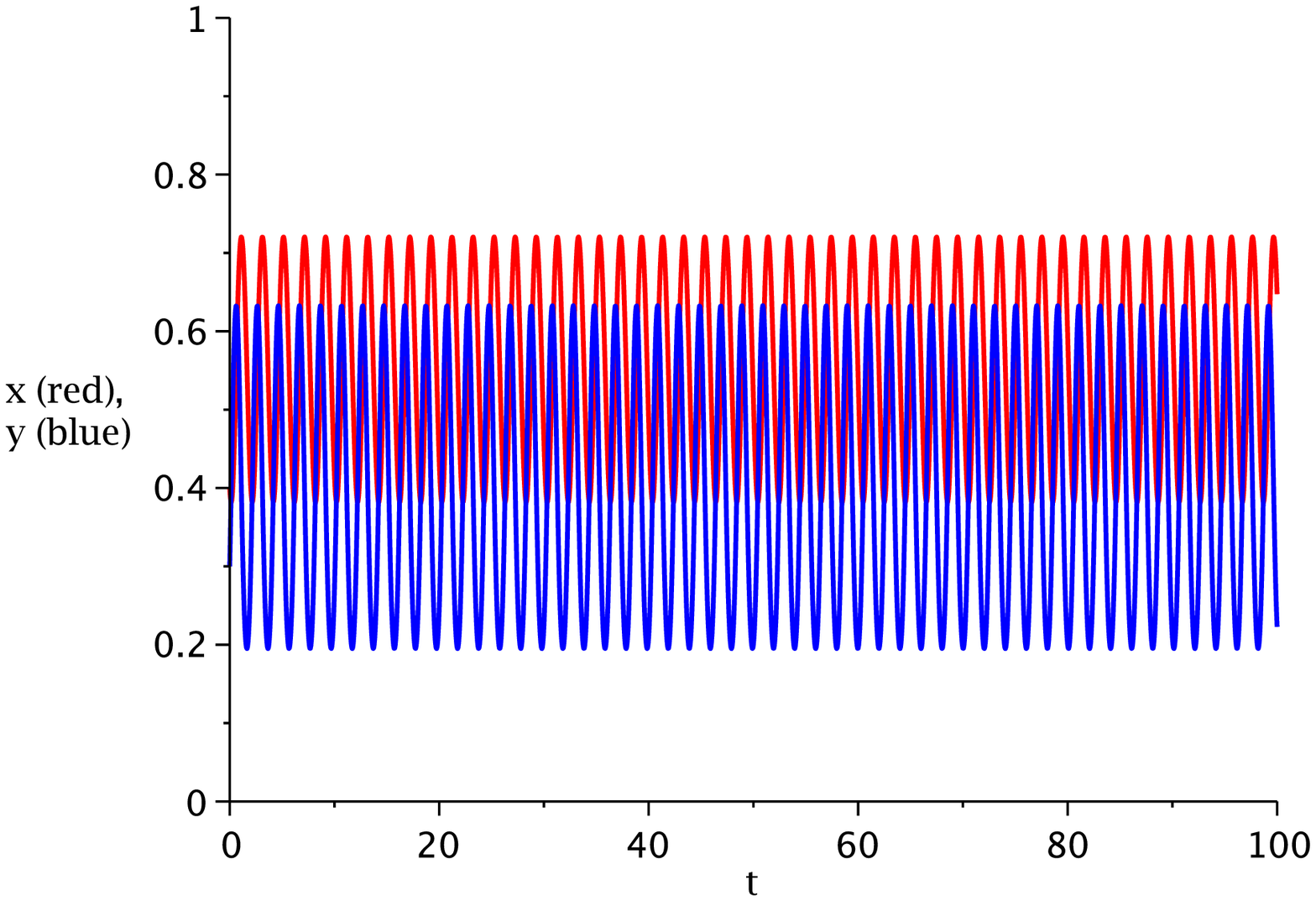,height=6cm,width=9cm,angle=0}
	\end{tabular}
	\vspace{-.5cm}
	\caption{Phase diagram (left) displaying the replicator dynamics vector field and one particular trajectory (thick line) corresponding to the initial conditions $(x_0=0.4,y_0=0.3)$ and evolution of $x$ (red) and $y$ (blue) over time (right) for the same initial conditions $(x_0=0.4,y_0=0.3)$.}
	\label{fig:1}
\end{figure}

Thus, in the payoff matrix, the $F$-type seller is able to recover part of his losses in each trade that takes place.\footnote{for example, avoiding purchase cancellation or avoiding a perishable good reaching expiration date.} Such recovered margin is modelled by a positive value of $\mu$ on top of a negative $\psi$ when adapting the e-commerce payoff matrix in (\ref{Matrix-I}) for addressing the traditional retail sector during a COVID-19 lockdown. We still keep our e-commerce market assumption in which trading only takes place when the game strategy profiles $(E,F)$ or $(L,S)$ are played. In our retail market, an $F$-type seller differs from the $S$-type seller due to having a more just in time logistic chain. Once a consumer makes a purchase in the physical store, the $F$-type seller produces the good and makes it available for in store collection faster than an $S$-type seller. Despite selling homogeneous goods, the faster seller charges a higher price due to such differentiation in product availability. Given our adaptation of assumptions for using the e-commerce market game in the context of the traditional retail market, the consumers' utility ordering in (\ref{Matrix-I}) is still preserved. In order to simulate such market context, we assumed the following parameters: $U_1=8$, $U_2=5$, $U_3=2$, $U_4=1$, $\psi=-10$ and $\mu=2$. The game payoff matrix becomes:
\begin{equation*}
	\bordermatrix {&F&S\cr
		E &8,-8& 1,0 \cr
		L &2,0& 5,-10\cr} 
\end{equation*}
and the RD system given by (\ref{RDp}) and (\ref{RDq}) becomes:
\begin{eqnarray*}
\dot{x}&=&x(1-x)\left[ 10y-4\right]\\
\dot{y}&=&y(1-y)\left[-18x+10\right]
\end{eqnarray*} 
The results of the numerical simulation are displayed in Figure \ref{fig:1}. The phase space displays closed clockwise orbits about the interior stationary point, leading to an oscillatory pattern in both populations of sellers and buyers over time. Starting at initial conditions $x_0=0.4$ and $y_0=0.3$, initially 40\% of the population of buyers is eager to shop even during pandemic restrictions and 30\% of the population of sellers offers the possibility of in store pre-arranged collection. From the phase diagram trajectory on the left panel of Figure \ref{fig:1}, initially the share $y$ of sellers offering collection increases, leading to an increase in $x$ over time, i.e., the share of eager consumers rises. Such increase discourages sellers to offer the service, leading to a fall in $y$ followed by a fall in $x$. The cycle then restarts and continues forever as shown on the right panel of Figure \ref{fig:1}.
The numerical simulation shows that none of the two types of sellers is able to take over the supply side of the market and the same happens with regard to consumption behaviour. None of the consumers' types is able to take over the demand side of the market. Under negative margins $\psi<0$, the market conditions did not enable (faster) in store collection practices (here a proxy for delivery) to become a standard in the traditional retail sector during the COVID-19 pandemic. Actually, the pre-pandemic practices were kept, with a mixed market trend in which some sellers do and others do not offer delivery. Such a result found in the numerical simulation for particular values of the model parameters is actually a general result, which we state and prove below.

\textbf{Proposition 1:} Whenever $\psi<\psi+\mu<0$ in the payoff matrix (\ref{Matrix-I}), the proportions of each type of agent in each population keep oscillating over time, without converging to an evolutionary equilibrium in the long run.

\textbf{Proof.} From the Jacobian matrix in (\ref{Jacobiana}), the eigenvalues computed at each corner stationary point are $\lambda_1(0,0)=\frac{\partial \dot{x}}{\partial x}=U_4-U_2<0$; $\lambda_1(1,1)=U_3-U_1<0$; $\lambda_1(0,1)=U_1-U_3>0$; $\lambda_1(1,0)=U_2-U_4>0$ and $\lambda_2(0,0)=\frac{\partial \dot{y}}{\partial y}=-\psi\lessgtr 0$; $\lambda_2(1,1)=-\psi-\mu\lessgtr 0$; $\lambda_2(0,1)=+\psi\gtrless 0$; $\lambda_2(1,0)=+\psi+\mu\gtrless 0$. Thus, depending on the sign of $\psi$, $(0,0)$ and $(1,1)$ are either attractors of the vector field in the phase diagram or saddle points. In contrast, $(0,1)$ and $(1,0)$ are either sources repelling the vector field or saddle points. The eigenvalues of the interior stationary point are given by $\lambda_{1,2}=\pm\sqrt{\frac{\partial \dot{x}}{\partial y}\cdot\frac{\partial \dot{y}}{\partial x}}$, thus, depending on the sign of $\psi$, we get either pure imaginary eigenvalues $\left(\psi<0\Rightarrow\frac{\partial \dot{y}}{\partial x}<0\right)$ or real eigenvalues with opposite sign $\left(\psi>0\Rightarrow\frac{\partial \dot{y}}{\partial x}>0\right)$.   

Given that $\psi<0$ and $\mid\psi\mid>\mid\mu\mid$, the eigenvalues of $\Omega^*$ computed at the corner stationary points are real numbers with opposite signs, thus each corner stationary point is a saddle point of the phase space, attracting the vector field in one direction and repelling it in the other direction, see Figure \ref{fig:1} (left). On the other hand, $\Omega^*$ has pure imaginary eigenvalues when those are computed at the interior stationary point. In the latter case, the non-linear ODE system given by (\ref{RDp}) and (\ref{RDq}) is non-hyperbolic at the interior stationary point. Thus, with regard to the system stability, one cannot assume that the non-linear system will behave as the linearised system in the neighbourhood of the stationary point (Hartman-Grobman theorem). Despite this, a well known result in evolutionary game theory (see Hofbauer and Sigmund, 1998) shows that the interior stationary point of our replicator dynamics system with pure imaginary eigenvalues is a centre and the vector field in the phase space displays closed orbits about that point. In the time-domain, such orbits translate into an oscillatory behaviour of the shares of each type of agent in the population and the game has no evolutionary equilibrium in the long run$\bullet$

\subsection{E-commerce Sector}
\label{subsec:2}
\begin{figure}[H]
	\centering
	\begin{tabular}{ccc}
		\epsfig{file=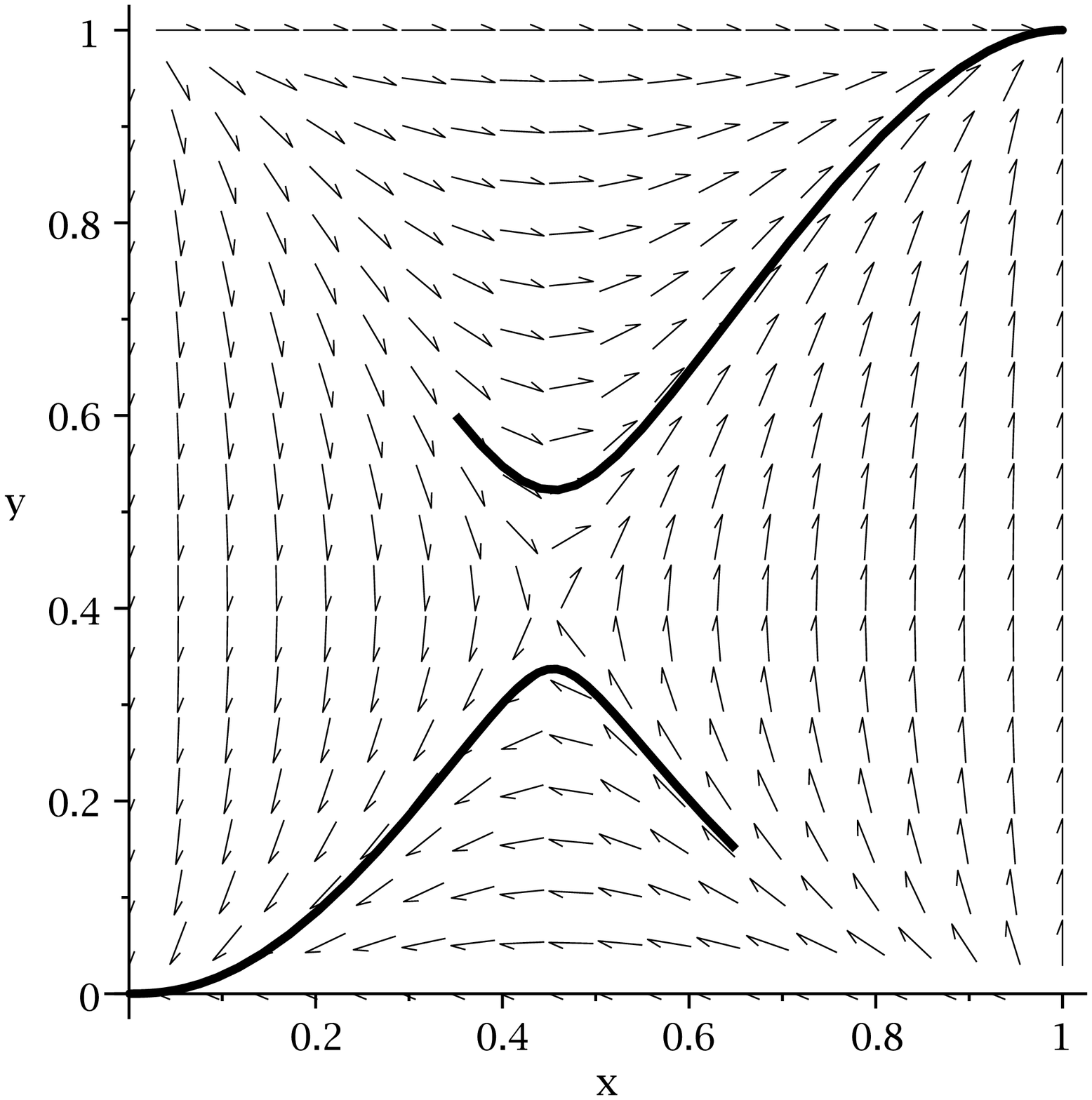,height=5cm,width=5cm,angle=0}&
		\epsfig{file=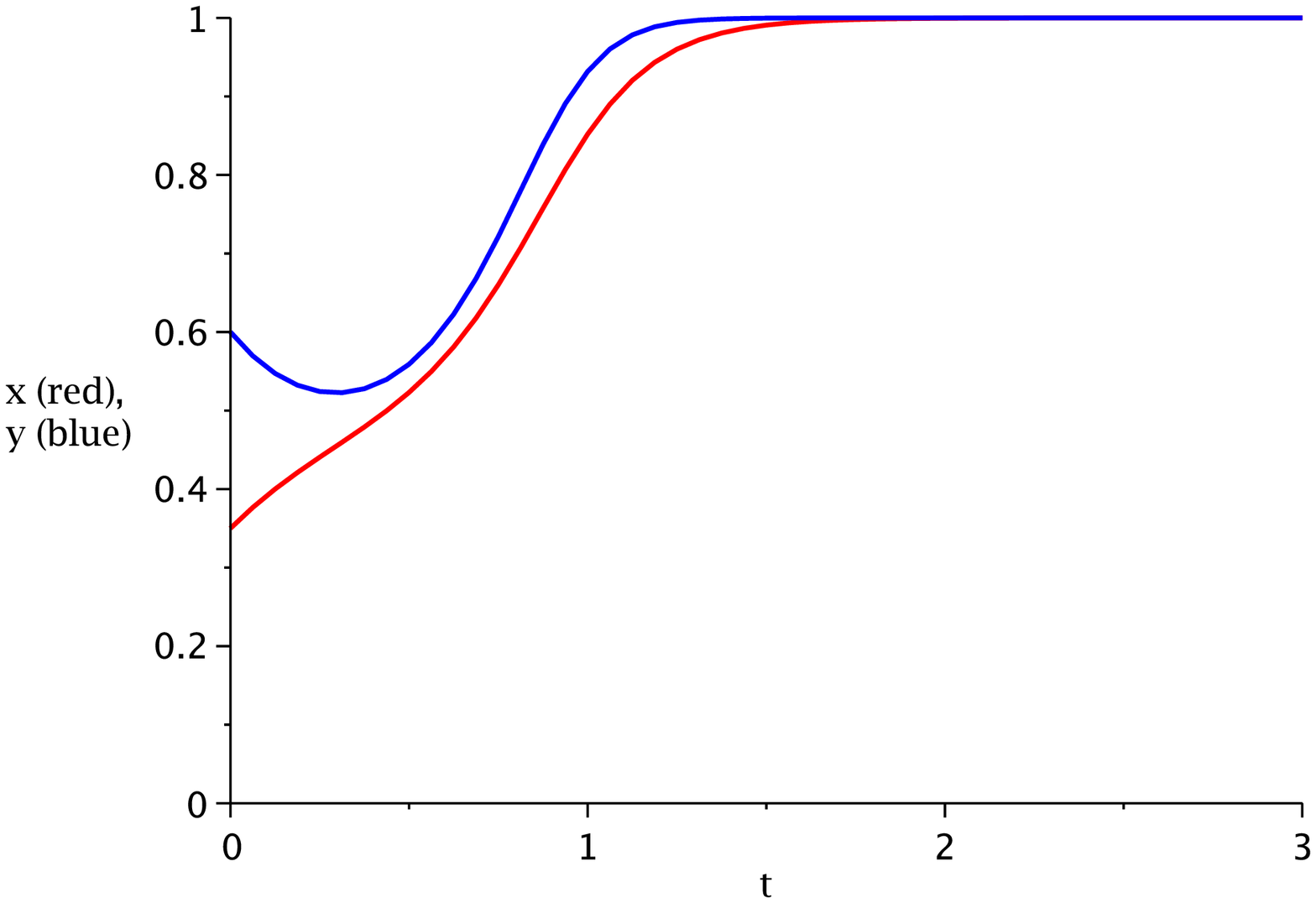,height=5cm,width=5.25cm,angle=0}&
		\epsfig{file=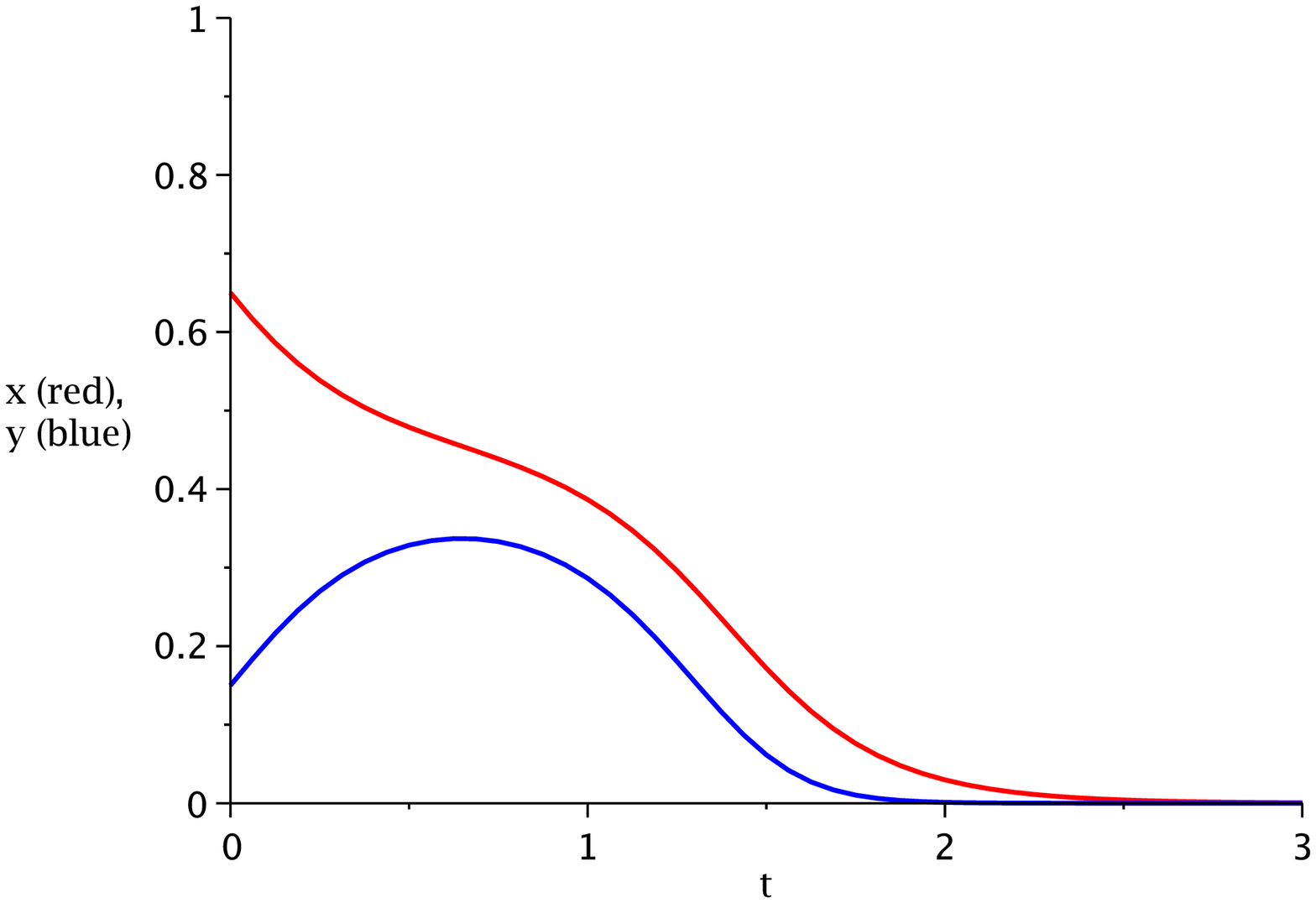,height=5cm,width=5.25cm,angle=0}
	\end{tabular}
	\vspace{-.5cm}
	\caption{Phase diagram (left panel) displaying the RD vector field and two particular trajectories (thick lines) corresponding to the initial conditions $(x_0=0.65,y_0=0.15)$ and $(x_0=0.35,y_0=0.60)$; and evolution of $x$ (red) and $y$ (blue) over time: ($x_0=0.35,y_0=0.60$, centre), ($x_0=0.65,y_0=0.15$, right).}
	\label{fig:2}
\end{figure}

The second case studied is the application of our model to the e-commerce business. In contrast to the traditional retail business, during the COVID-19 pandemic, e-commerce experienced a sharp increase in both demand and revenue, leading to operating profits better than those anticipated by the market consensus for the sector, i.e., $\psi>0$. In such context, improvements in delivery policies and logistics network became a must in order to be a differentiated seller and grow market share. Regarding e-commerce, we carried out a numerical simulation with the same parameters of the first simulation but with positive operating margin ($\psi=10$). The game payoff matrix is:
\begin{equation*}
	\bordermatrix {&F&S\cr
		E &8,12& 1,0 \cr
		L &2,0& 5,10\cr} 
\end{equation*}
And the corresponding RD system given by (\ref{RDp}) and (\ref{RDq}) becomes:
\begin{eqnarray*}
\dot{x}&=&x(1-x)\left[10y-4\right]\\
\dot{y}&=&y(1-y)\left[22x-10\right] 
\end{eqnarray*}
The phase diagram and the evolution of both populations over time corresponding to the simulation results are displayed in Figure \ref{fig:2}.

With producers facing positive operating markup, the resulting game phase diagram in the left panel of Figure \ref{fig:2} displays a very different topology when compared to that of Figure \ref{fig:1} in subsection \ref{subsec:1}. One can see that there are two evolutionary equilibria located at the corner points $(0,0)$ and $(1,1)$ and two basins of attraction in the interior of the phase space leading to each equilibrium. When the evolutionary dynamics converges to the equilibrium located at $(0,0)$, the eager type consumer and the seller with fast logistics become extinct in their respective populations over time. The most interesting equilibrium for society regarding trading efficiency is the one located at $(1,1)$ in which all consumers look for fast delivery and producers are all committed with holding an efficient logistics network.
In the central panel of Figure \ref{fig:2}, starting at the initial conditions with an e-commerce market with a demand population composed of 35\% of eager consumers and a supply population composed of 60\% of sellers with a fast delivery logistics, there are two opposing forces in the market. From the payoff matrix, on the one hand, eager type consumers have an incentive to grow their share of participation in the demand side of the e-commerce market given that, for a majority of $F$-type sellers, their expected utility is higher than the expected utility of an $L$-type consumer, i.e., $\bar{U}_E=8\cdot 0.60+1\cdot 0.40>2\cdot 0.60+5\cdot 0.40=\bar{U}_L$. On the other hand, $F$-type sellers initially start to decrease their share in the supply side due to the small share of consumers looking for their differentiated logistics, i.e., $\bar{\Pi}_F=12\cdot 0.35+0\cdot 0.65<0\cdot 0.35+10\cdot 0.65=\bar{\Pi}_S$. But, at some point, with the growth of the $E$-type consumer share, such trend switches in the opposite direction and fast delivery corporate culture starts to grow among sellers. In the long run, the e-commerce market is dominated by such trend among sellers as well as dominated by consumers that are eager to receive their goods as soon as possible. Both the $L$ and the $S$ types of agents disappear from the e-commerce market. A market with a large supply of efficient delivery and logistic services is able to drive consumers' taste toward an increasing demand for those services.

In contrast, the e-commerce market conditions could be very different. Assuming initial conditions $x_0=0.65,y_0=0.15$, the right panel in Figure \ref{fig:2} displays convergence to $(0,0)$ in the long run. In such case, despite having almost 2/3 of the consumption market composed of eager consumers looking for an efficient and fast delivery logistics, the share of producers meeting such corporate strategy was too low, with only 15\% of $F$-type sellers. As a result, over time, eager consumers decreased their share in the demand side of the market. Despite the share of sellers reacting to this consumption trend, even with the share of sellers adopting fast delivery growing over time, such a strategic move was not strong enough to change the trend in the consumption market. Eager consumers continued to decrease monotonically their share in the population until becoming extinct. With such trend, $F$-type sellers also faced a negative impact and, after reaching almost 40\% of the share of the producers' population, they also started to decrease their share until disappearing from the population. In the long run, an e-commerce market dominated by old standards, i.e., consumers who are happy to wait longer for their products and firms delivering their goods through a slow network, becomes the norm. The result found in the numerical simulation for particular values of the model parameters is a general one, which we state and prove below.

\textbf{Proposition 2:} Whenever $\psi+\mu>\psi>0$ in the payoff matrix (\ref{Matrix-I}), the game between sellers and consumers has two evolutionary equilibria and the population dynamics converges to either $(0,0)$ or $(1,1)$ in the long run, depending on the location of the initial conditions of both populations with regard to the basins of attraction in the phase diagram. 

\textbf{Proof.} From the Proof of Proposition 1, given that $\mid\psi\mid>\mid\mu\mid$, but instead we have $\psi>0$, the eigenvalues of $\Omega^*$ computed at the corner stationary points are both negative real numbers at $(0,0)$ and $(1,1)$, thus both are asymptotically stable points and evolutionary equilibria of the game. In contrast, the eigenvalues are both positive real numbers at the corners $(0,1)$ and $(1,0)$, thus unstable points. On the other hand, $\Omega^*$ has real eigenvalues with opposite signs when those are computed at the interior stationary point. The latter is a saddle point and, together with the unstable corner points, those three points define the separatrix between each basin of attraction of the evolutionary equilibria $(0,0)$ and $(1,1)$, see Figure \ref{fig:2}. The population dynamics will converge to one of those equilibria depending on the location of the initial conditions in the phase space $\bullet$

The phase diagram in Figure \ref{fig:2} explains the important effort that e-commerce firms deserved to their logistics services and delivery network during the COVID-19 pandemic. With social restrictions, there was an important increase in the demand of goods purchased via online marketplaces. This is similar to the initial condition corresponding to the trajectory leading to the equilibrium at $(0,0)$ in Figure \ref{fig:2}. In line with such initial condition, well experienced e-commerce firms such as Amazon, Mercado Livre and Alibaba, despite having already a well developed and tested delivery network, were not prepared to deal with such an increase in demand. An example in Brazil were e-commerce companies that still relied too much on the Brazilian regular postal company ECT. Consumer complaints regarding parcels being late or even lost rocketed at the beginning of the pandemic. In the following months, e-commerce companies moved in the direction of improving and developing their own network of warehouses in order to make delivery faster and reliable. In Figure \ref{fig:2} (left), this is mathematically modelled as a vertical move in the phase diagram. For given proportion of eager consumers $x$, sellers invest in order to raise the share of $F$-type sellers and avoid a fall in the share of eager consumers willing to buy from them. This corporate action moves the populations' state coordinate in the phase diagram from the basin of attraction leading to $(0,0)$ in the long run into the other basin of attraction leading to the evolutionary equilibrium $(1,1)$.

Empirical evidence of the delivery bottleneck described above, which was followed by investment in logistics to overcome delivery and distribution constraints, can be easily found. Taking the Brazilian case as an example, during the second quarter of 2020, the number of consumer complaints regarding online purchases increased 53.5\% when compared to the same period one year before (Nalin, 2020). One of the main reasons for those complaints was the failure to meet the delivery deadline agreed during the purchase process. The majority of the Brazilian e-commerce sellers were not prepared for the peak in demand that took place in the beginning of the pandemic. On top of this, sellers and consumers had to deal with a strike carried out by the state owned Brazilian postal service company. In order to bypass such bottlenecks in the delivery logistic chain, sellers made important investments in new warehouses and improved the use of other transportation and courier companies. In September 2020, Amazon announced the opening of its fifth and largest warehouse in Brazil in order to meet the increase in demand due to social isolation measures (Alves, 2020). According to Amazon, the logistic facility would help to meet a larger consumer demand and make deliveries faster. In October 2020, Amazon announced a plan to make further investments in its logistic chain and new warehouses in Brazil. 
Regarding other large players, Mercado Livre announced the hiring of 60 additional trucks to help with the delivery of goods, while Via Varejo concluded the acquisition of ASAPlog, a start-up with logistic expertise in making the connection between transportation companies within long distance delivery operations (Alves, 2020).  

\section{Conclusion}
\label{sec:conclusion}
We developed an evolutionary game in order to model the interplay between consumers and producers when trade takes place on an e-commerce marketplace. The game payoff matrix was then adapted to analyse trading patterns developed during the COVID-19 pandemic in both the traditional retail sector and the e-commerce sector. While in the latter massive investment in delivery logistics and warehouses allowed for the establishment of a trend in which fast delivery and eager consumers are the norm, in the former sector high street shops remained in a sort of non-technological trap in which traditional street shoppers and inefficient companies did not disappear.

In a moment when mass vaccination becomes a reality and the world appears to start to recover after the worst impact of the pandemic, it becomes evident that the traditional retail sector not only seems to be doomed to disappear in the long run, but such trend accelerated during the pandemic in favour of the e-commerce sector. Even in very profitable sectors such as the Brazilian banking sector, such trend seems to have accelerated with more customers and savers opening accounts in online banks and fintechs, in detriment of traditional banks with several retail physical branches. As an example, looking at the close share prices of the main Brazilian banks trading in the B3 stock market on 06-May-21, while the yield to date share price variation was -0.29\%, -11.65\%, -12.41\% and -22.29\% for Bradesco (BBDC4), Ita\'u (ITUB4), Santander (SANB11) and Banco do Brasil (BBAS3), respectively, the variation for Banco Inter (BIDI11) and BTG Pactual (BPAC11) shares was +117,05\% and +21.62\%, respectively for the same period.  

\section*{Acknowledgements}
This research was supported by CNPq, research grant no. 307437/2019-1.

\end{document}